# Multi-stage Attack Detection and Prediction Using Graph Neural Networks: An IoT Feasibility Study

Hamdi Friji[*‡], Ioannis Mavromatis[†], Adrian Sanchez-Mompo[†], Pietro Carnelli[†], Alexis Olivereau[*], and Aftab Khan[†]
[*]CEA, LIST, Communicating Systems Laboratory, F-91191 Gif-sur-Yvette, France
[†]Bristol Research and Innovation Laboratory, Toshiba Europe Ltd., Bristol, UK
[‡]SAMOVAR, Télécom SudParis, Polytechnic Institute of Paris, 91120 Palaiseau, France
Email: hamdi.friji@cea.fr

*Abstract*—With the ever-increasing reliance on digital networks for various aspects of modern life, ensuring their security has become a critical challenge. Intrusion Detection Systems play a crucial role in ensuring network security, actively identifying and mitigating malicious behaviours. However, the relentless advancement of cyber-threats has rendered traditional/classical approaches insufficient in addressing the sophistication and complexity of attacks. This paper proposes a novel 3-stage intrusion detection system inspired by a simplified version of the Lockheed Martin cyber kill chain to detect advanced multi-step attacks. The proposed approach consists of three models, each responsible for detecting a group of attacks with common characteristics. The detection outcome of the first two stages is used to conduct a feasibility study on the possibility of predicting attacks in the third stage. Using the ToN IoT dataset, we achieved an average of 94% F1-Score among different stages, outperforming the benchmark approaches based on Random-forest model. Finally, we comment on the feasibility of this approach to be integrated in a real-world system and propose various possible future work.

## I. INTRODUCTION

Nowadays, network security is becoming crucial, with information systems and networks constantly at risk of cyber-attacks [1]. Intrusion Detection Systems (IDSs) are pivotal in defending these systems through providing early alerts of potential security breaches. By effectively detecting and mitigating intrusions, companies can monitor the risks associated with cyber-threats and protect confidential data from unauthorised access, data breaches, and financial losses [2]. Cyber-attacks can lead to severe consequences, including disruption of services, compromise of personal information, and damage to an organisation's reputation. The impact of such incidents can be far-reaching, affecting not only the targeted entity but also its partners and customers. Consequently, it is paramount to find solutions to detect attacks and, in the best-case scenario, to predict some attacks.

In industry, we find three main types of IDSs [3], i.e.: Network Intrusion Detection Systems (NIDSs), Host-based IDSs, and Hybrid IDSs, classified based on their functionality and monitoring approach. This work presents an NIDS approach where the network is monitored by collecting packets, creating communication flows, and extracting features that characterise them. Such features are the DNS-, SSL-, HTTP-activities, and statistical features (e.g. number of source/destination packets, etc.). In this work, we consider only the NIDSs approach where we can find two main sub-categories in the state-of-the-art solutions:

- Anomaly-based NIDS [2]: Anomaly-based NIDS learns a baseline of normal behaviour/activities and detect any deviation of the learned normal behaviour as a potential attack. Anomaly-based IDSs suffer from a high false positive rate and fail to detect sophisticated attacks that mimic normal behaviour patterns or exhibit subtle deviations.
- Signature-based NIDS [2]: also known as rule-based NIDS, consists mainly of using a database of pre-defined attack signatures to distinguish between normal and malicious flows. Signature-based NIDS effectively detect well-known attacks but may struggle with detecting new or unknown threats [2].

The classical intrusion detection approaches [4] are context-agnostic, and they consider only the flows' features to distinguish between normal and malicious communications. Moreover, the diversity of attacks types in the networks makes the learning of Machine Learning (ML) models harsh. In most cases, it leads to a high false positive rate, making the models unusable in practical applications [3]. From another point of view, having multiple models, each specialising in detecting one class, allows for fast detection at the cost of energy/time for training all models.

Most successful real-life attacks are composed of several steps. Each step has a specific goal and is performed using specific techniques, making it difficult for classical intrusion detection frameworks to detect complex attacks (i.e., multi-step attacks).

Having a model that considers the temporal evolution of an attack has several advantages in practice and it is highly appreciated by cyber-engineers. In industry, we can find frameworks that describe the evolution of complex attacks, namely CKC chains [5] (e.g., Lockheed Martin's and MITRE ATT&CK frameworks), that are intended to facilitate the defending and monitoring of the attack by the cybersecurity engineers. The knowledge acquired from these frameworks could be used to improve IDSs.

In order to overcome the previously mentioned limitations/challenges, we propose a novel multi-stage IDS, specifically, a 3-stage IDS. Our proposed framework leverages a simplified version of the Lockheed Martin's Cyber Kill Chain

(CKC) [5] in order to detect complex attacks as they are evolving. The detection is performed in a context-aware/agnostic approach using ML and Graph Neural Network (GNN) algorithms [6]. The first two stages, namely Reconnaissance and Privilege Escalation (PE), generate alerts and embeddings as output. The embeddings are used, in a second step, for predicting the users that will be targeted by a third stage's attacks, namely Access Exploitation (AE).

The remainder of this paper is structured as follows: related work is reviewed in Section II. Section III introduces our proposed approach. Section IV covers dataset selection and investigation. Section V presents our results and discusses the challenges we faced. Finally, Section VI concludes the paper and highlights potential future research work.

## II. RELATED WORK

Prior IDS/cybersecurity research mainly focused on using a single ML for binary or multi-class classification. For example, [4], [7], [8] show several approaches that use ML and Deep Learning (DL) models to classify normal flows using their features only. Despite their ability to detect multiple attack types, they often struggle to detect advanced attacks (e.g., Advanced Persistent Threads) and are easily affected by evading techniques (e.g., IP Spoofing).

Recent advancements in the field of complex attack detection have leveraged multi-step models as a promising approach. For example, the authors of [9] suggest a distributed intrusion detection system tailored for IoT environments. The primary objective of this IDS is to detect different forms of cyber-attacks effectively. To achieve this, the system operates in three steps: (1) categorising device types, (2) detecting malicious network flows, and (3) identifying the specific types of attacks being carried out. This system does not include any context in its detection and does not provide any practical advantages.

To overcome the limitations introduced by single model approaches, authors of [10], presented an approach for detecting web-based attacks using a multi-model framework. The authors suggested combining multiple detection models to improve the accuracy and effectiveness of intrusion detection. The proposed approach exploits several models, such as rule-based systems, anomaly detection, and ML algorithms, to analyse web traffic and predict potential attack patterns. Similarly too the previous works, this system ignores the context of communications and doesn't provide a way to follow the evolving of complex attacks.

For the attacks prediction model, we rarely found works that investigated attack prediction due to the arduous nature of the task. For example, the work in [11] suggest an approach for predicting cyber-attacks using DL algorithms. Additionally, authors in [12] also proposed a prediction model that exploits Bayesian networks to predict cyber-attack. The Bayesian networks represent the relationships between different variables and their influence on the attacks, and they discussed the construction and training of the Bayesian network using historical attack data.

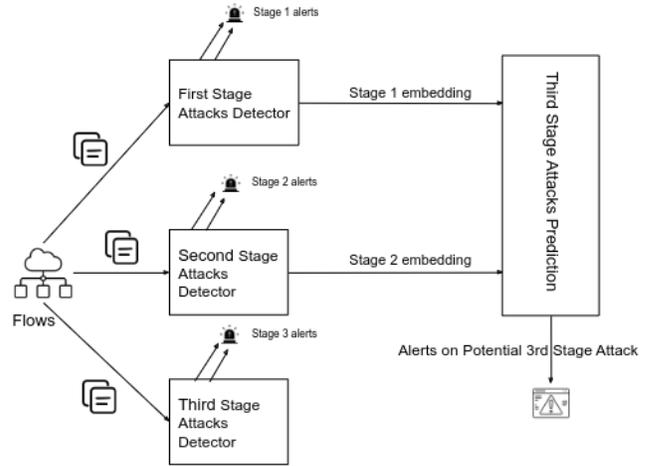

Fig. 1. Illustration of the Different Components of our Proposed Framework

To the best of our knowledge, no work performed intrusion detection using any CKC (i.e. Lockheed Martin's) informed architecture, and there is no previous research about predicting attack stages from other stages of the same multi-step attack.

## III. MULTI-STAGE IDS

In this section, we present our proposed 3-stage IDS, and we introduce the CKC, specifically the Lockheed Martin's CKC. We, also, explain how our IDS inspires from the CKC to detect complex attack while they are evolving.

### A. The Proposed Framework

This work develops and evaluates the effectiveness of detecting intrusions using three specialised models. The approach, exhibited in Fig. 1, consists of three ML/Graph Neural Network (GNN) models inspired by the Lockheed Martin's CKC, namely stage detectors and defined in Section III-D, working in parallel. Each responsible for detecting a specific stage of the 3-stage CKC defined in Section III-C. In other words, each model will perform binary classification where the class labelled as "Stage $i$" is for the attacks included in the investigated stage "$i$", and the second class labelled as "Other" includes the other attack's stages and the normal flows.

Each model is trained using a version of the dataset that was labelled to perform binary classification where the first class is the "Other" class (i.e., Normal and other malicious flows) and the second class is stage $i$ attack, where $i \in \{1, 2, 3\}$).

Consequently, during the system's functioning, the cybersecurity engineers receive three types of alerts from three different components (i.e., stages). Each alert aims to notify of a potential attack from a specific stage. The engineers will have more knowledge about the possible attack type and will promptly determine the best techniques to stop the attacks.

Having three stages provides the Security Operations Center team with more information about the detected attack type, and hence, it facilitates the task of determining the techniques to monitor this attack. On the other side, having only three stages reduces the detection's complexity compared to multi-attacks detection (i.e., single model responsible for detecting all the attacks). Accordingly, having track of the attacks' stages facilitates the scenario re-creation by the forensics team after incidents. For example, they can group the alerts per target and obtain all the attacks performed against that target from each stage, and hence, they can determine all the users who were potentially included in the attack.

The stage detectors are composed of a context-aware/agnostic models that perform the classification of the flows. The context-aware model, based on GNN algorithms, is responsible for performing flows classification and generating an embedding. Moreover, the embeddings generated by stages 1 and 2 context-aware models' (i.e., the embeddings were generated only for the detected malicious flows by stage 1 and 2 detectors) are used by another component in the proposed framework to predict potential stages 3 attacks.

### B. The Cyber Kill Chain

CKC [5] is a framework that describes the stages of a cyber-attack from the perspective of the attacker. It enables understanding of the cyber-attack lifecycle and helps cybersecurity engineers identify and mitigate potential threats. One of the most known CKC is Lockheed Martin's CKC that differs from other CKC [5] by describing the different stages of a cyber-attack as follows :

*1) Reconnaissance:* In this stage, the attacker gathers information about the target network or system. This can involve passive activities like Open-Source Intelligence (OSINT) gathering or active scanning and probing to identify potential vulnerabilities and targets.

*2) Weaponization:* The attacker develops or acquires the tools, exploits, or malware necessary to carry out the attack. This stage involves crafting or customising the attack payload to deliver it effectively.

*3) Delivery:* The attacker delivers the weaponized payload to the targeted system(s). This can be done through various means, such as email attachments, malicious links, or compromised websites.

*4) Exploitation:* In this stage, the attacker exploits vulnerabilities in the target system to gain unauthorised access or execute malicious code. This can involve leveraging software vulnerabilities, misconfigurations, or social engineering techniques to compromise the target.

*5) Installation:* Once the attacker has gained a foothold in the target system, they proceed to install persistent mechanisms to maintain access and control. This can include deploying backdoors, remote access tools, or creating user accounts to ensure continued access.

*6) Command and Control (C2):* The attacker establishes communication channels or C2 infrastructure to remotely manage/control the compromised systems. This allows them to execute commands, exfiltrate data, or launch further attacks.

*7) Actions on Objectives:* In this final stage, the attacker achieves their primary objectives, which could involve data exfiltration, system disruption, privilege escalation, or any other malicious activities aligned with their goals.

Lockheed Martin's CKC is highly exploited in the cybersecurity industry. It offers a structured framework for understanding the stages of a cyber-attack, enabling organisations to identify vulnerabilities and develop effective defence strategies. The CKC facilitates better alignment of defences, resource allocation, and implementation of countermeasures. It also fosters communication and collaboration among security professionals. Overall, the adoption of the CKC has greatly improved the industry's ability to detect, respond to, and mitigate cyber-threats effectively [5].

### C. 3-stage Cyber Kill Chain

In our approach, we aim to propose an IDS inspired by the previously mentioned CKC's. Hence, we group the six detectable stages of the Lockheed Martin's CKC into a three main stages as cited in the sequential:

*1) Stage 1 Reconnaissance Stage:* Mainly, it represents the Reconnaissance attack, where the attacker gathers information about the target. The information gathering step is essential in any serious attacks because the attackers need the information to understand the targeted environment and be able to exploit any found Vulnerabilities.

*2) Stage 2 Privilege Escalation (PE):* It combines the Delivery and Exploitation phases, where the intruder exploits the gathered information, the vulnerabilities in the system and security flaws found during Stage 1 attacks (i.e. Reconnaissance attacks) to gain unauthorised access to the target system.

*3) Stage 3 Access Exploitation (AE):* This stage includes the Installation, Command and Control, and Actions on Objectives attacks. It aims mainly to cause harm to the system after acquiring the necessary privileges from stage 2.

The Weaponization phase is not included in the simplified CKC since the attacker is not interacting with the system during this phase.

AE stage contains the most harmful attacks; hence, detecting, mitigating, or stopping it early is important. The two other stages are equally important in the attack-kill chain but they are less aggressive towards the network and users.

The CKC framework and the simplified version consider the temporal aspect of cyber-attacks. While the stages provide a sequential progression of an attack, they also acknowledge that attacks may not always follow a linear timeline. The temporal aspect recognises that attacks can occur over an extended period and may involve iterative or simultaneous activities across different stages.

### D. Stage's Attack Detectors

In each stage, a stacking of two models is used to recognise malicious flows and determine their stages. The models architectures considers the flows interconnection with other flows in the network to create a context of the communication and enhance the effectiveness of differentiating between normal and malicious flows. The stage detectors' architectures are shown in Fig. 2. The flows are conveyed through the following two models:

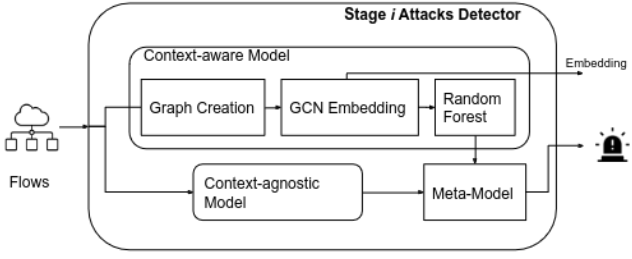

Fig. 2. Stage Detectors Architecture

- Context-aware Model: This model transforms the flows into a graph where two flows are connected if they are generated or received by the same user [13]. This graph structure provides the detection more robustness against evading techniques (i.e., IP spoofing) as explained in [13]. The graph is, afterwards, forwarded to a Graph Convolutional Network (GCN) that will generate an embedding. The GCN are considered an excellent data embedders that can extract relevant topological information from the graph structure. The embedding includes information about users' interactions and creates a context for the investigated communication. This embedding aims to improve the detection accuracy by incorporating contextual information, capturing dynamic relationships, and adapting to changing network environments. Moreover, GCNs can be retrained on new datasets to include new attacks/features. The mathematical formulation of GCNs, which explains their ability to extract topological information, is detailed later in this section III-D1.
- Context-agnostic Model: This model considers the flow's features without any consideration of any interaction with other users. In our work, this is a random-forest based model trained and tested directly on processed raw data.

The previous two models are stacked using a Meta-model (i.e., logistic regression) that aims to combine the information coming from the context and the flow's features by ensuring a good balance between the two types of information. The stacking technique in ML and its advantages are explained in subsection III-D2.

*1) GCN Embedding:* GCNs are a type of neural network designed for graph-structured data. They are highly effective at learning node embeddings and capturing the topological pattern in the graph structure.

Let consider $\mathcal{G} = (\mathcal{V}, \mathcal{E})$ be an undirected graph, where $\mathcal{V}$ represents the set of nodes and $\mathcal{E}$ represents the set of edges. The graph can be represented as an adjacency matrix $\mathbf{A} \in \mathbb{R}^{N \times N}$, where $N$ is the number of nodes (i.e., flows) in the graph. Each element $A_{ij}$ of the adjacency matrix indicates the presence or absence of an edge between nodes $i$ and $j$. Additionally, let $\mathbf{X} \in \mathbb{R}^{N \times F}$ be the feature matrix, where $F$ is the number of input features for each node (i.e., the number of attributes assigned to each flow).

The graph convolutional layer [6] is the key component of GCNs. It aggregates information from a node's neighbourhood and updates the node's embedding accordingly. Let $\mathbf{H}^{(l)} \in \mathbb{R}^{N \times D}$ be the node embeddings at layer $l$, where $D$ is the dimensionality of the embeddings.

The update rule for the $i$-th node's embedding at layer $l+1$ can be defined as:

$$\mathbf{H}_i^{(l+1)} = \sigma \left( \sum_{j \in \mathcal{N}(i)} \frac{1}{\sqrt{|\mathcal{N}(i)||\mathcal{N}(j)|}} \mathbf{H}_j^{(l)} \mathbf{W}^{(l)} \right),$$

where $\sigma$ is an activation function (e.g., ReLU), $\mathcal{N}(i)$ represents the set of neighbouring nodes of node $i$, and $\mathbf{W}^{(l)} \in \mathbb{R}^{D \times D}$ is a learnable weight matrix at layer $l$. The normalisation factor $\frac{1}{\sqrt{|\mathcal{N}(i)||\mathcal{N}(j)|}}$ adjusts the aggregation weights to account for different neighbourhood sizes.

A GCN consists of multiple graph convolutional layers that takes as input the graph, followed by a final layer that produces the node embeddings. The node embeddings generated by the Reconnaissance and Privilege Escalation stages will be further used for the Access exploitation attacks prediction.

*2) Machine Learning Models Stacking:* Model stacking is an ensemble learning technique where multiple models are combined hierarchically to enhance predictive performance [14]. Base models $f_i$ are trained on the training dataset $X_{\text{train}}$ to generate predictions $y_{\text{base}}^i = f_i(X_{\text{train}})$. These predictions serve as inputs for a higher-level model $g$ that learns to make final predictions $\hat{y}$. The stacking ensemble is trained by minimising the discrepancy between the predicted values and the true target variable. During prediction, the trained ensemble uses the base models to generate predictions for new input data $X_{\text{test}}$, which are then combined by the higher-level model to obtain the final predicted target variable $\hat{y}_{\text{test}} = g(f_1(X_{\text{test}}), f_2(X_{\text{test}}), \ldots, f_N(X_{\text{test}}))$. Model stacking allows the ensemble to leverage the diverse strengths of individual models (i.e., context-aware/agnostic models) and improve overall predictive effectiveness.

### E. Access Exploitation Attacks Prediction

The Access Exploitation attacks, such as Ransomware attacks, Backdoor attacks, are the most harmful attacks, and thus it is really important to investigate solutions that provide possibilities to predict or early-stop them.

This subsection checks the possibility of predicting AE stage attacks. We aim to study the feasibility of predicting if the Reconnaissance attack followed by the privilege escalation attack (i.e., stage 2 attack) were both successful and the attacker gained the required access to harm the system through an AE attack (e.g., ransomware attack). We also check if the prediction could be performed using the embeddings extracted from the first- and second-stage attacks.

Predicting the third stage attack is challenging since each stage of a complex attack could be done using different IPs (i.e., the attacker frequently changes its IP address to evade detection), or a whole team could perform the attack. Thus, we are not predicting which attackers will perform the third-stage attacks. Instead, we are predicting the possibility that a specific user will be targeted by an AE attack. Afterwards, we can use the first and second stages' alerts to determine the included attackers. In other words, we predict if a specific host (e.g., server or normal user) will be targeted by a third-stage attack after receiving Reconnaissance and PE attacks. Being alerted of a potential attack on a specific machine/user is beneficial in anticipating the attack by taking proactive steps.

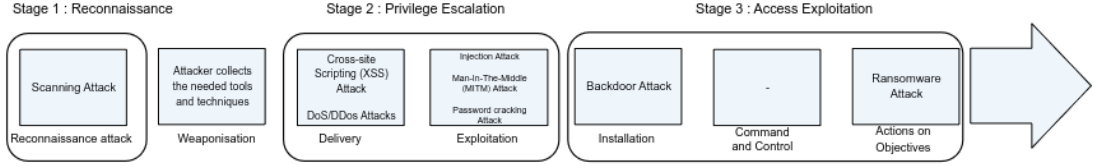

Fig. 3. The Simplification of the Lockheed Martin's Cyber Kill Chain.

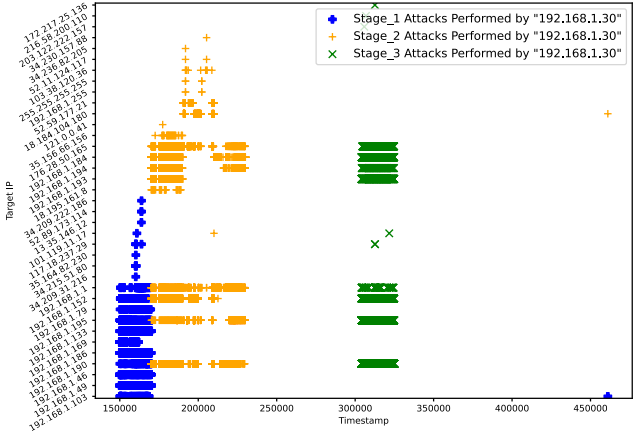

Fig. 4. ToN IoT: Visualization of the Attack's Stages Performed by Attacker with IP Address "192.168.1.30"

For the training and testing of AE attacks, another dataset based on the embedding that allows performing prediction per targeted user should be prepared to be fed to a multi-input Recurrent Neural Network (RNN) [15].

Multi-Input RNNs are a variant of RNNs that can handle multiple input sequences simultaneously. They are particularly useful for step-wise problems, where each input represents a different step of a process.

The architecture of a multi-input RNN extends the basic RNN model by incorporating multiple input sequences. In our case, we have two input sequences denoted as $X_1$ and $X_2$, composed of $T$ component, each one characterise each of the first two stages, and it is defined as $X_1 = \left[\mathbf{x}_1^{(1)}, \mathbf{x}_2^{(1)}, \ldots, \mathbf{x}_T^{(1)}\right]$, and $X_2 = \left[\mathbf{x}_1^{(2)}, \mathbf{x}_2^{(2)}, \ldots, \mathbf{x}_T^{(2)}\right]$, where $\mathbf{x}_t^{(1)}$ and $\mathbf{x}_t^{(2)}$ represent the $t$th element in the first and second input sequences, respectively.

Multi-input RNNs aim to exploit the temporal features present in each stage of the input sequences. By processing the input vectors at each time step, the RNN captures the dependencies within each stage and combines information from all stages in the hidden state $\mathbf{h}_t$. This enables the model to understand the temporal patterns and make accurate stage-wise predictions.

The RNN can be represented as follows:

$$\mathbf{h}_t = \text{activation}\left(\mathbf{W}_{hx1}\mathbf{x}_t^{(1)} + \mathbf{W}_{hx2}\mathbf{x}_t^{(2)} + \mathbf{W}_{hh}\mathbf{h}_{t-1} + \mathbf{b}_h\right),$$

where $\mathbf{h}_t$ is the hidden state of the RNN at time $t$. " activation" is the activation function used in the RNN cell (e.g. tanh and ReLu). The $\mathbf{W}_{hx1}$ and $\mathbf{W}_{hx2}$ are the weight matrices that map the input sequences $X_1$ and $X_2$ to the hidden state, respectively. The $\mathbf{W}_{hh}$ is the weight matrix for the recurrent connections, which maps the previous hidden state $\mathbf{h}_{t-1}$ to the current hidden state $\mathbf{h}_t$, and $\mathbf{b}_h$ is the bias term for the hidden state. After processing both input sequences, the final hidden state $\mathbf{h}_t$ is be passed through a fully connected layer with softmax activation to generate class probabilities $\mathbf{y}_t = \text{softmax}(\mathbf{W}_{yh}\mathbf{h}_t + \mathbf{b}_y)$, where $\mathbf{y}_t$ is the output at time $t$. $\mathbf{W}_{yh}$ is the weight matrix that maps the hidden state to the output, and $\mathbf{b}_y$ is the bias term for the output.

## IV. DATASET INVESTIGATION

For the Intrusion Detection task, several available datasets [3] can be used to train/test frameworks responsible for attack detection using binary/multi-class classification. However, there is no dataset specialised in performing multi-stage detection and attacks' prediction.

Several datasets were assessed/investigated, such as CI-CIDS2017, Edge IIoT, WUSTL-IIoT-2021, but none of the them satisfied the criteria to train/test this work. However, the ToN IoT [16] allowed an effective training of GNNs, and provided the required information to create the three sub-datasets to train/test the stage detectors.

Fig. 3 maps ToN IoT dataset attacks to Lockheed Martin's CKC stages. We illustrate the grouping of stages for constructing the 3-stage CKC. Notably, the dataset lacks instances targeting the "Command and Control" stage, as evident in Figure 3.

Most attacks in the ToN IoT dataset respect Lockheed Martin's CKC, hence the 3-stage CKC. Fig. 4 plots the users in the ToN IoT who were targeted by the attacker with IP address "192.168.1.30" (i.e., Y-axis) and its occurrence during the evolution of time, represented by the Timestamp (i.e., X-axis). The figure shows that the attacker has followed the 3-stage CKC in most of his attacks.

From another point of view, Fig. 5 shows the temporal succession of malicious flow stages generated by the attackers (i.e., in the Y-axis ) that have targeted the user with the "101.119.11.11" IP address.

We acknowledge that in a real-life scenario, attackers may not always stick with the succession used in Lockheed Martin's CKC, and we can have more random behaviours. Also, stages may be repeated before passing to the next stage in the CKC, but in all cases, the attacks should perform the Reconnaissance attack at a certain time to gather information about the target; hence, the first stage in the 3-stage CKC is respected. Afterwards, the attacker will perform either a Delivery or Exploitation attack; hence the PE stage is respected. The repetitive behaviour of the first two stages of our CKC does not affect the system since an alert will be generated for each behaviour, and the cyber-engineers will notice it.

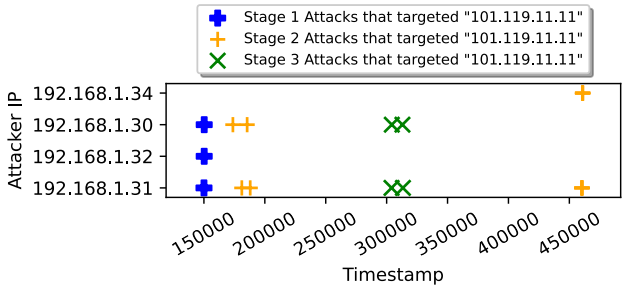

Fig. 5. ToN IoT: Visualisation of the Attack's Stages that Targeted the User/Machine with IP Address "101.119.11.11"

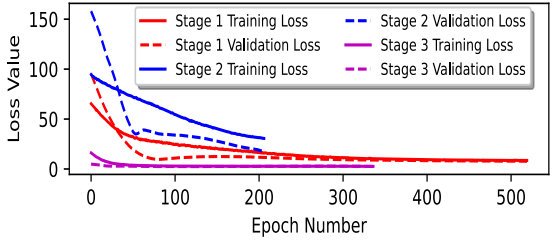

Fig. 6. CrossEntropy Loss Evolution for the Three Stage Detectors during Training and Testing

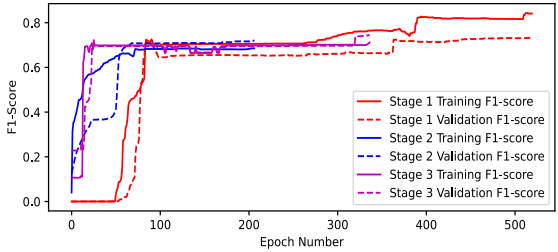

Fig. 7. F1-score Evolution for the Three Stage Detectors during Training and Testing

To ensure a reliable evaluation, we have used an IP-splitting evaluation technique to assess the performance of the stage detectors. This technique consists of selecting different users for the training and testing. It has several advantages, such as avoiding any data leakage between training and testing due to the similarity between flows of the same user, and it allows to check how good the model generalises on unseen attacks. Finally, we decided to remove the $Src\_Ip\_Bytes$ features since, when combined with other features, it allows a superficial separation between malicious and normal flows, which is not the case in the real-life data.

## V. Results & Discussion

In this section, we present the obtained results of our framework using the ToN IoT dataset. Afterwards, we discuss the challenges and how it could be improved in future work.

### A. Stages' Attacks Detector Evaluation

Figs. 6 and 7 provide insights into the performance of the used models for each stage. Curves in Figs. 6 illustrates the evolution of the training and validation binary crossEntropy loss for each model over time, while Fig. 7 showcases the evolution of the F1-score [17] for the same models.

The trends and patterns in the first set of curves, included in Fig 7, show that during the first epochs, the models are not generalising well on the validation set. During the evolution of training, the models gets more and more accurate in their predictions, and they converges at the end of training, which confirms the learning's effectiveness and stability.

Moreover, the evolution of the F1-score during training and testing conveys the models' ability to correctly classify instances of both positive and negative classes. By analysing the F1-score curves, we can notice that the models' performance is increasing during the training, which reveals that the model is learning and enhancing its abilities to generalise and accurately classify unseen data, which is shown in the increase of the F1-score in the validation loss curve.

The provided confusion matrices in Figs. 8 to 10 offer an insight into the classification performance of the models for the "Normal", "Stage 1", "Stage 2", and "Stage 3" categories. In the three models, the "Normal" category exhibits a high rate of correct classifications as "Other". This proves that the models effectively recognise instances belonging to the "Normal" category and assign them to the appropriate class (i.e., "Other" class).

The confusion matrices for stage detectors in Figures 8 to 10 demonstrate that the Stage 1 model achieves a high rate of correct classifications of the "Normal" class (i.e., 0 misclassification for "Stage 1" detector and 39, 120 for "Stage 2" and "Stage 3" respectively). Moreover, each model is highly accurate in detecting the investigated stage (i.e. 8 misclassification as "Other" for "Stage 1" detector and 373, 340 for "Stage 2", "Stage 3" respectively). Furthermore, there are some misclassifications of stage 3 flows as stage 2, and stage 2 as stage 3, indicating challenges in correctly identifying and differentiating between these two categories. However, from a binary classification perspective, the confusion between stage 2 and stage 3 is less important than misclassification with normal flows.

In Table I, we endorse the previous graphical results with a numerical evaluation using several metrics [17] (i.e., F1-score, Precision, Recall, and False Positive Rate (FPR)). Since no previous work investigated the same problem using a similar approach, we recreate a Random-Forest-based system proposed in [18] to use it as a benchmark for our work. Each of the proposed stage detectors outperformed the benchmark models by almost 3% in the F1-score. The stage 1 detector exhibited the highest overall performance among the three models, and it achieved an F1-score of 0.995 in comparison to 0.976. Stage 2 and stage 3 detectors reached, respectively, 0.93 and 0.893, in comparison to the benchmark models that reached, 0.905 and 0.864. We notice for each model a balance between precision and recall that conveys the effectiveness of the learning. Accordingly, the stage detectors reduce the FPR [17] to achieve 0.07, 0.134, and 0.207, respectively for stage 1, stage 2, and stage 3. The other the benchmark models attended a higher FPR for all stages.

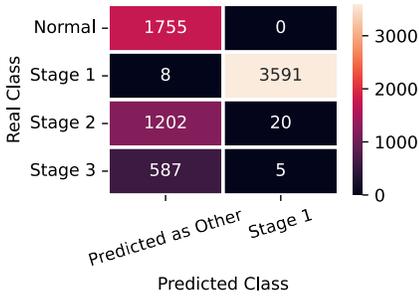

Fig. 8. Stage 1 Detector's Confusion Matrix

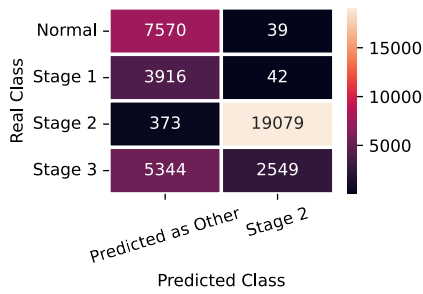

Fig. 9. Stage 2 Detector's Confusion Matrix

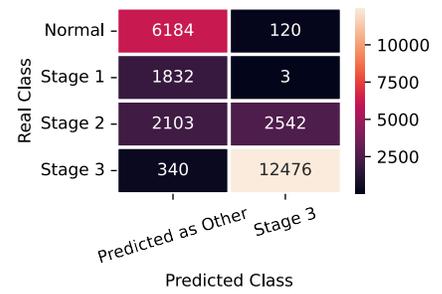

Fig. 10. Stage 3 Detector's Confusion Matrix

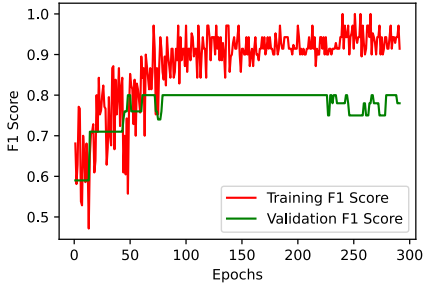

Fig. 11. F1-score Evolution

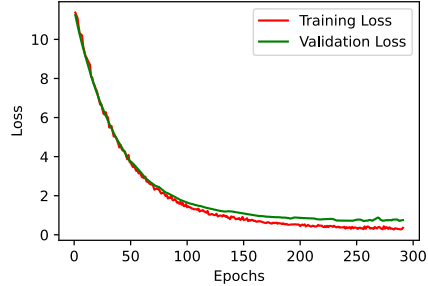

Fig. 12. CrossEntropy Loss Evolution

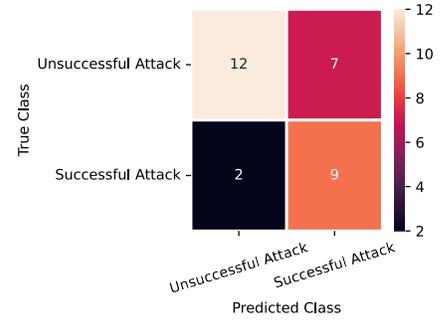

Fig. 13. AE Prediction Confusion Matrix

TABLE I
NUMERICAL VALIDATION OF STAGE DETECTORS

| Metrics | F1-score | Precision | Recall | FPR |
|---|---|---|---|---|
| Stage 1 Detector | 0.995 | 0.993 | 0.998 | 0.007 |
| Benchmark Model 1 | 0.976 | 0.978 | 0.974 | 0.021 |
| Stage 2 Detector | 0.930 | 0.88 | 0.980 | 0.134 |
| Benchmark Model 2 | 0.905 | 0.882 | 0.930 | 0.123 |
| Stage 3 Detector | 0.893 | 0.824 | 0.973 | 0.207 |
| Benchmark Model 3 | 0.864 | 0.808 | 0.929 | 0.220 |

These results shows that the stage 3 models is not as effective as the first two stages. This could be explained by the difficulty of detecting the AE attacks due to their complexity.

To summarise, the context-aware detection in stage detectors, explained in III-D, allows to improves the detection performance of the three models and reduces the FPR.

*1) Access Exploitation Attacks Prediction:* The Access Exploitation attack's prediction consists in predicting if a user is going to be targeted by an AE attack after receiving a Reconnaissance and Privilege escalation attack. For that purpose a labelled dataset should be created from the first two stage detectors' output and the ToN IoT dataset. Unfortunately, the obtained dataset was small due to the low of number of users in the initial testbed. For each user (i.e., targeted user) in the obtained dataset, we have two lists of embeddings, an IP address, and a label. The two lists contain the embedding coming from the stage detectors, and that will be used as input for the RNN. The label "0" denotes if the two first attacks were not followed by an AE attack and "1" for the opposite case.

To overcome the limitations posed by the reduced size of the dataset, we adapted the exploited RNN architecture in order to obtain stable training and to check if the model is really able to predict the AE attacks from the first two stages. On the other hands, the set of samples in which a user has received stage 1 and 2 attacks but not stage 3 attacks is small, which makes the problem even harder.

The Figures 11 and 12 show the evolution of the F1-score and the CrossEntropy loss during the training and testing. We can notice instabilities in the calculation of the F1-score, and this is due to the small number of samples since a single right/wrong prediction will change the F1-score considerably. However, the overall trend of the F1-score and the loss shows that the model is learning and he is getting more and more capable of predicting if the user/machine is going to be targeted by an AE attack.

The Fig. 13 shows the confusion matrix of the RNN. The model is making mistakes in its prediction, and this could be explained by the lack of data. However, learning from a poor quality dataset with tiny and unbalanced number of samples and being able to distinguish between some of the successful and unsuccessful attacks is highly promising and encouraging Further Inquiry.

To summarise, despite the errors made by the model, the classification results and the training stability evince that the prediction performance could be enhanced more by improving the quality of the created dataset. They, also, demonstrate that it is highly possible to predict when an Access Exploitation attack follows a Reconnaissance and Privilege Escalation attack. These results can be improved more to ensure a more accurate prediction, and that will ensure the anticipation of the most harmful attacks. The possible improvement of this section is discussed in the next paragraph.

*B. Challenges & Possible Improvements*

Our framework's results demonstrate promising outcomes and advantages in its detection and prediction capabilities, including improved accuracy, reduced false positives, adaptability to emerging threats, and comprehensive visibility. Additionally, it empowers cyber-engineers with practical information to facilitate their tasks in monitoring and facing attacks. However, to obtain a reliable system, several challenges must be addressed.

A potential improvements of this work is to create a specialised dataset that enables CKC-based multi-stage intrusion detection, contains more attacks/attackers, and provides a balance distribution between attacks that will/will not be followed by AE attacks. The datasets will assist in enhancing the training of the whole system and specifically the training and the evaluation of the AE attacks prediction. Another considerable challenge is the windowing strategy of the analysed flows on each prediction. In other words, how many flows should we consider at each functioning of the system since the system may suffer to detect very long attacks unless we increase the windowing size and hence increase complexity.

Another possible improvement of the systems efficiency is to distributively create the graph used in the context-aware detection, and this could be performed in the edge context to ensure a lower training and/or inference time, and hence the solution could be used for real-time prediction.

## VI. Conclusion

This work proposes a novel IDS approach that allows the detection of advanced multi-step attacks and attacks' scenario recreation. The proposed approach is composed of 3 stages, namely Reconnaissance, Privilege Escalation, and Access Exploitation. Each one of the three stages includes a group of attacks that have similar objectives. For each stage, we have a specialised model, namely stage detector, trained to detect the attacks included in the investigated stage. The stage detectors are composed of two sub-detectors, context-aware/agnostic detectors. Each stage detector sends alerts to the cyber-engineers and generates an embedding from the detected attack data that will be used later to predict how potentially a users will be targeted by the third stage attack (Access Exploitation attack). The results show that by including the context-aware detection, we are able to improve the detection F1-score by around 3% in comparison to the benchmark models attending an average of 94% among three stages. The AE attacks prediction results confirm the feasibility of predicting the happening of AE attacks against some users who have already been targeted by Reconnaissance and PE attacks.

## VII. Acknowledgement

This work was supported in part by Toshiba Europe Ltd. and in part by the GreenEdge project (grant no. 953775, European Union's Horizon 2020 Marie Skłodowska Curie Innovative Training Network).